\documentclass[preprintnumbers,showpacs,amsmath,amssymb,floatfix,prd,twocolumn,
superscriptaddress,nofootinbib]{revtex4}
\usepackage{latexsym}
\usepackage{epsfig}
\usepackage{amssymb}
\begin{document}

\title{Noncommutative geometry inspired $3$-dimensional charged black hole solution in
 an anti-de Sitter background space-time}

\author{Farook Rahaman}
\email{rahaman@iucaa.ernet.in} \affiliation{Department of
Mathematics, Jadavpur University, Kolkata 700032, West Bengal,
India.}

\author{Piyali Bhar}
\email{piyalibhar90@gmail.com } \affiliation{Department of
Mathematics, Jadavpur University, Kolkata 700032, West Bengal,
India.}

\author{Ranjan Sharma}
\email{rsharma@iucaa.ernet.in} \affiliation{Department of
Physics, P. D. Women's College, Jalpaiguri 735101, India}

\author{Rishi Kumar Tiwari}
\email{rishitiwari59@rediffmail.com} \affiliation{Department of
Mathematics, Govt. Model Science College, Rewa, M.P., India.}

\begin{abstract}
We report a $3$-D charged black hole solution in an anti-de Sitter space inspired by
noncommutative geometry. In this construction, the black hole exhibits two horizons
 which turn into a single horizon in the extreme case. We investigate the impacts of
 electromagnetic field on the location of the event horizon, mass and thermodynamic
 properties such as Hawking temperature, entropy and heat capacity of the black hole.
  The geodesics of the charged black hole are also analyzed.
\end{abstract}

\maketitle

\section{Introduction}
In recent years, noncommutative geometry has received a considerable interest to get
insight into many conceptual issues in the quantum domain of high energy physics.
 It is argued that at sufficiently high energy scales the target-space structure
  coordinates may become noncommuting operators on a D-brane\cite{Witten,Seiberg}.
  Consequently, the noncommutativity of the space-time can be encoded in a commutator
   $\left[x^{\mu},x^{\nu}\right] = i \theta^{\mu \nu}$, where, $\theta^{\mu \nu}$ is
    an anti-symmetric matrix of dimension $[\text{length}]^{2}$ which determines the
     fundamental cell discretization of the space-time. It is similar to the way the
     Plank's constant $\hbar$ discretizes phase-space\cite{Smailagic}. The noncommutative
      approach provides the construction of a black hole with a minimum scale $\sqrt{\theta}$,
       known as noncommutative black hole\cite{Rizzo,Nicolini1}, whose commutative limit is
        the Schwarzschild metric. It is note worthy that noncommutative geometry is an
         intrinsic property of the space-time and it does not depend on the curvature
         of space-time.

A large number of studies inspired by noncommutative geometry are available in the
literature. Myung {\em et al}\cite{Myung} have discussed thermodynamic behaviour and
evaporation processes involved in a noncommutative black hole. In their investigations,
they have considered a noncommutative $3$-dimensional black hole whose commutative limit
 is the non-rotating Ba$\tilde{n}$ados, Teitelboim and Zanelli\cite{BTZ} (henceforth BTZ)
 black hole. Nozari and Mehdipour\cite{Nozari} have investigated the Hawking radiation
  from a noncommutative black hole. Rahaman {\em et al}\cite{Rahaman1}, in the context
  of galactic rotation curves, have shown that a noncommutative geometrical background
  is sufficient for the existence of a stable circular orbit and one need not consider
  dark matter in this formalism. In a separate paper, Rahaman {\em et al}\cite{Rahaman2}
   have shown that wormhole solutions exist up to $5$-dimensional space-times; however,
    they do not exist in dimensions $> 5$. Banerjee {\em et al}\cite{Banerjee} have
     investigated the thermodynamic properties (Hawking temperature, entropy and area law,
      in particular) of a Schwarzschild black hole in noncommutative space-times. By invoking
       a graphical analysis, the authors have shown the non-trivial behavior of the
        noncommutative Schwarzschild metric at different length scales. It has been
        shown that in the limit of large $r$ ($r >> \sqrt{\theta}$) outside the horizon,
         the metric behaves like the standard Schwarzschild solution, whereas in the limit
          of small $r$ ($r << \sqrt{\theta}$) inside the horizon, the space-time is
           de-Sitter type with a constant positive curvature.

\textbf{The motivation of the present work is to construct a
$3$-dimensional charged black hole solution based on
noncommutative geometry in an anti-de Sitter background
space-time. Using the commutator of operators corresponding to
space-time coordinates, i.e. (the integer D below is even), one
can encode the   non-commutativity of space-time $[ x^\mu,x^\nu]
= i \theta^{\mu \nu}$ . Here the  anti symmetric matrix
$\theta^{\mu \nu}$ assumes the block-diagonal form as\\
$\theta^{\mu \nu} = diag (\theta_1, ..............,
\theta_{D/2})$ where}
\[
        \theta_i=
            \left[ {\begin{array}{cc}
             0 &1 \\
              -1 &  0 \\
                               \end{array} }\right]
~~~for~ all,~ i=1,2,...,D/2
        \]
\textbf{ The fundamental  discretization  of  spacetime can be
determined by this antisymmetric  matrix  $\theta^{\mu \nu}$. For
flat spacetimes, it  is also recognized  that noncommutativity
removes the  point-like structures in favor of smeared objects.
 The fundamental  discretization  of  spacetime can be determined
by this antisymmetric  matrix  $\theta^{\mu \nu}$.   For flat
spacetimes, it  is also recognized  that noncommutativity removes
the  point-like structures in favor of smeared objects. One can
mathematically implement  the  effect of the smearing   with a
substitution of the Dirac-delta function by a Gaussian
distribution of minimal length $\sqrt{\theta }$. In particular,
here, we will consider  the energy density of a static and
spherically symmetric, smeared and particle-like gravitational
source.  Since we assumed  this particular  form  of the energy
density $\rho$ of the fluid ( which is smeared and particle-like
gravitational source and will be given in eq. (15) in the next
section  ), therefore, we call the solutions as ' Noncommutative
geometry inspired'.}

It is to be stressed here that gravitational analyses in ($2+1$)-dimensions have become
extremely fascinating with the discovery of the black hole solution by Ba$\tilde{n}$ados,
Teitelboim and Zanelli\cite{BTZ}. The BTZ metric describes the exterior gravitational
 field of a circularly symmetric Einstein-Maxwell system in the presence of a negative
 cosmological constant and is characterized by its mass, angular momentum and charge.
  In standard ($3+1$)-dimensions, Einstein-Maxwell systems for static spherically
   symmetric charged distributions of matter have investigated by many (see for example,
Tikekar and Singh\cite{Tikekar}, Sharma {\em et
al}\cite{Sharma1}, Thirukkanesh and
     Maharaj\cite{Thirukkanesh} and also references therein). Analytic models of static
     charged fluid spheres has been reviewed by Ivanov\cite{Ivanov}. In ($2+1$)-dimensions,
      Rahaman {\em et al}\cite{Rahaman3} have developed a model for a charged gravastar
      in an anti-de Sitter background space-time. Corresponding to the static BTZ exterior,
       Cataldo {\em et al}\cite{Cataldo} have investigated the properties of a circularly
        symmetric static charged fluid distribution. Larra$\tilde{n}$aga and
         Tejeiro\cite{Larranaga} have obtained a charged black hole solution in
          $3$-D anti-de Sitter space inspired by noncommutative geometry where a
           Gaussian distribution of charge has been considered. In an earlier work,
            Rahaman {\em et al}\cite{Rahaman4} have studied BTZ black holes inspired
            by noncommutative  geometry. The current investigation is an extension of
             Ref.~\cite{Rahaman4} where we have incorporated a distribution of charge
              into the system. The resultant charged black hole appears to have two
               horizons which, however, get reduced to a single horizon in the extreme
               case. Thermodynamic properties such as Hawking temperature, entropy and
               heat capacity of the charged black hole have been analyzed. Effective
                potential and behaviour of test particles around the black hole have
                also been outlined.
\\
\\

\section{Interior space-time}
We write the line element for the interior space-time of a static circularly symmetric
 charged distribution of matter in ($2+1$)-dimensions in the form
\begin{equation}
ds^{2} = -e^{\nu(r)}dt^{2} + e^{\lambda(r)}dr^{2}+r^{2}d\phi^{2}.\label{eq1}
\end{equation}
The energy-momentum tensor corresponding to the charged fluid distribution is assumed to be
\begin{eqnarray}
T_{ij}^{\text{Total}} &=& (\rho+p_r)u_iu_j - p_r g_{ij} +(p_t-p_r)\eta_{i}\eta_{j}\nonumber\\
 &&-\frac{1}{4\pi}\left(F_a^{c}F_{bc}-\frac{1}{4}g_{ab}F_{cd}F^{cd}\right),\label{eq2}
\end{eqnarray}
where, $\rho$ is the energy-density, $p_r$ is the radial pressure and $p_t$ is the
 tangential pressure. In (\ref{eq2}), $u^i$ is the velocity of the fluid and $\eta^i$
 is a radial unit vector satisfying the relation $u^i u_i = -\eta^i \eta_j =1$. The
 electromagnetic field tensor is governed by the Maxwell's equations
\begin{equation}
F^{ij}_{;j}=-4\pi J^{i},\label{eq3}
\end{equation}
where, $ J^{\mu}$ is the current $3$-vector defined as
\begin{equation}
J^{i} = \sigma(r)u^{i}.\label{eq4}
\end{equation}
In (\ref{eq4}), $\sigma(r)$ is the proper charged density of the distribution.
 The electromagnetic field tensor may be expressed as
\begin{equation}
F_{ij} = E(r)\left(\delta_i^{t}\delta_j^{r}-\delta_i^{r}\delta_j^{t}\right),\label{eq5}
\end{equation}
where, $E(r)$ is the electric field intensity. In the presence of a negative cosmological
 constant ($\Lambda<0$), the Einstein-Maxwell equations
\begin{equation}
R_{ij}-\frac{1}{2}Rg_{ij}+\Lambda g_{ij} = -2\pi T_{ij}^{\text{Total}},\label{eq6}
\end{equation}
for the metric (\ref{eq1}) yield the following set of four independent equations
(in geometrized units $G = c = 1$)
\begin{eqnarray}
\frac{\lambda'e^{-\lambda}}{2r} &=& 2\pi \rho+E^{2}+\Lambda,\label{eq7}\\
\frac{\nu'e^{-\lambda}}{2r} &=& 2\pi p_r -E^{2}-\Lambda,\label{eq8}
\end{eqnarray}
\begin{equation}
\frac{e^{-\lambda}}{2}\left(\frac{\nu'^{2}}{2}+\nu''-\frac{1}{2}\nu'\lambda'\right)
 = 2\pi p_t +E^{2}-\Lambda,\label{eq9}
\end{equation}
\begin{equation}
\sigma(r)=\frac{e^{-\frac{\lambda}{2}}}{4\pi r}(rE)',\label{eq10}
\end{equation}
where a `prime' denotes differentiation with respect to the radial parameter $r$.
 Eq.~(\ref{eq10}) can equivalently be expressed as
\begin{equation}
E(r)=\frac{4\pi}{r}\int_0^{r}\tilde{r}\sigma(\tilde{r})e^{\frac{\lambda(\tilde{r})}
{2}}d\tilde{r} =\frac{q(r)}{r},\label{eq11}
\end{equation}
where $q(r) $ is the total charge contained within a sphere of radius $r$.

We assume that the volume charge density has the form
\begin{equation}
\sigma(r)e^{\frac{\lambda(r)}{2}} = \sigma_0r^{n},\label{eq12}
\end{equation}
where $n$ is an arbitrary constant and $\sigma_0$ is the central
charged density. \textbf{Though it is true that one is free to
choose an infinite number of possible ansatz, the main motivation
for a particular choice in our model was to generate solutions
which could be considered as physically meaningful. Secondly,
solutions obtained earlier for the neutral case can be easily
regained from our solution for the specified ansatz. This
enables  one to investigate the effects of electromagnetic field
on the system explicitly. The choice (12) of the charge density
is only dictated by the mathematical convenience   and has no
physical justification since it gives the energy divergent at
infinity. However, this particular choice is different from the
Gaussian distribution of charge assumed
 in Ref.~\cite{Larranaga}. } We will  solutions obtained earlier for the neutral
case can be easily regained from our solution for the specified
ansatz.

  Consequently, from Eq.~(\ref{eq11}) we get
\begin{eqnarray}
E(r)&=& \frac{4\pi \sigma_0}{n+2}r^{n+1},\label{eq13}\\
q(r) &=& \frac{4\pi \sigma_0}{n+2}r^{n+2}.\label{eq14}
\end{eqnarray}
Further, we consider the maximally localized source of energy of the static spherically
symmetric charge distribution as a Gaussian distribution with minimal width $\sqrt{\theta}$
 expressed as\cite{Rahaman4},
\begin{equation}
\rho = \frac{M}{4\pi \theta}e^{-\frac{r^2}{4\theta}},\label{eq15}
\end{equation}
where $M$ is the total mass of the source diffused throughout a region of linear dimension
 $\sqrt{\theta}$.

 Substituting (\ref{eq13}) and (\ref{eq15}) in (\ref{eq7}), we obtain
\begin{equation}
\frac{\lambda'e^{-\lambda}}{2r} = \frac{M}{2\theta}e^{-\frac{r^{2}}{4\theta}}+\left(\frac{4\pi
 \sigma_0}{n+2}  \right)^{2}r^{2n+2}+\Lambda,\label{eq16}
\end{equation}
which, on integration, yields
\begin{equation}
e^{-\lambda} = -A+2M e^{-\frac{r^{2}}{4\theta}}-\frac{1}{n+2}\left(\frac{4\pi \sigma_0}
{n+2}\right)^{2}r^{2n+4}-\Lambda r^{2},\label{eq17}
\end{equation}
where, $A$ is a constant of integration. In Ref.~\cite{Rahaman1} it has been shown that
the constant of integration $A$ plays the role of the mass of the black hole, i.e., $A = M$.
 In our construction, if we set $\sigma_0 = 0$ and consider the limit $\frac{r}{\sqrt{\theta}}
 \rightarrow \infty $, then it follows that $A = M$. Accordingly, we rewrite Eq.~(\ref{eq17}) as
\begin{equation}
e^{-\lambda} = -M+2M e^{-\frac{r^{2}}{4\theta}}-\frac{1}{n+2}\left(\frac{4\pi \sigma_0}
{n+2}\right)^{2}r^{2n+4}-\Lambda r^{2}.\label{eq18}
\end{equation}
Note that the vacuum Einstein field equations in ($2+1$)-dimensional space-time in the
 presence of a negative cosmological constant $(\Lambda<0)$ admit a
black hole solution known as BTZ\cite{BTZ} solution for which we have $g_{rr}=\left(g_{tt}
\right)^{-1}$. To retain the structure of BTZ black hole,
we must have $e^{\nu} = e^{-\lambda}$. From Eqs.~(\ref{eq7})-(\ref{eq8}), it then follows that
\begin{equation}
p_r = -\rho,\label{eq19}
\end{equation}
which is  the `$\rho$ vacuum' EOS in reference to `zero-point energy of quantum
fluctuations'\cite{Rahaman4}.

Combining Eqs.~(\ref{eq7})-(\ref{eq9}) and (\ref{eq19}), we obtain
\begin{equation}
\frac{e^{-\lambda}}{2}(\lambda'^{2}-\lambda'') = 2\pi p_t + E^{2}-\Lambda.\label{eq20}
\end{equation}
From Eq.~(\ref{eq16}), we get
\begin{eqnarray}
e^{-\lambda}(\lambda''-\lambda'^{2}) = \frac{M}{\theta}e^{-\frac{r^{2}}{4\theta}}\left(1-\frac{r^{2}}{2\theta}\right)\nonumber\\
+2(2n+3)\left(\frac{4\pi\sigma_0}{n+2} \right)^{2}r^{2n+2}+2\Lambda.\label{eq21}
\end{eqnarray}
Combining Eqs.~(\ref{eq20}) and (\ref{eq21}), we obtain
\begin{equation}
p_t = -\frac{M}{\pi \theta}e^{-\frac{r^{2}}{4\theta}}\left(1- \frac{r^{2}}{2\theta}\right)
-\frac{n+2}{\pi}\left(\frac{4\pi\sigma_0}{n+2}\right)^{2}r^{2n+2}.\label{eq22}
\end{equation}
We, thus, have a noncommutative geometry inspired analytic model of a $3$-D charged black
 hole. In this construction, the spacetime near the origin behaves as
\begin{eqnarray}
e^\nu  = e^{-\lambda} = -M+2M \left[ 1 - \frac{r^{2}}{4\theta} +
\frac{r^{4}}{32\theta^2} +  {\cal{O}}(r^8) \right] \nonumber\\
 -\frac{1}{n+2}\left(\frac{4\pi
\sigma_0}{n+2}\right)^{2}r^{2n+4}-\Lambda r^{2}.\label{eq23}
\end{eqnarray}
Note that for $n=0$, by adjusting the parameters suitably, one can regain the space-time
for the ($2+1$)-dimensional static charged distribution
\begin{eqnarray}
e^\nu  = e^{-\lambda} = M-\left[  \frac{M}{2\theta} + \Lambda
\right]  r^{2}
  \nonumber\\
 + \left[ \frac{2M}{32 \theta^2} - 2\pi^2 \sigma_0^2\right] r^{4},\label{eq24}
\end{eqnarray}
obtained earlier by Liang and Liu\cite{Liang}.

In the following sections, we shall analyze some features of our charged BTZ black hole.

\section{Features of the charged BTZ black hole}

\subsection{Formation of event horizons}
\begin{figure}
\centering
\includegraphics[scale=.4]{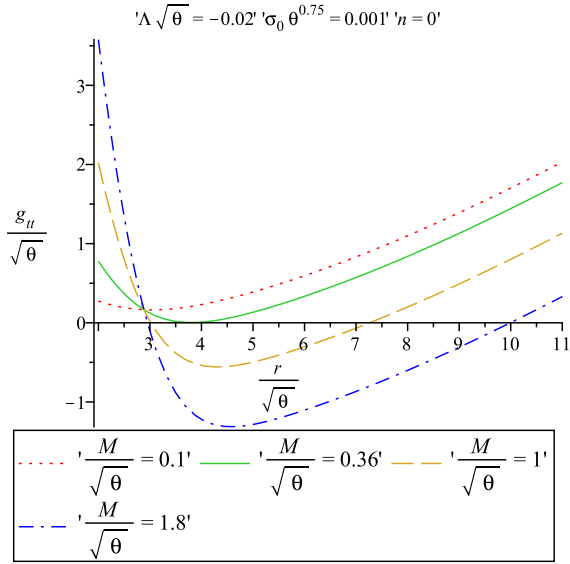}
\caption{$\frac{g_{tt}}{\sqrt{\theta}}$ plotted against $\frac{r}{\sqrt{\theta}}$.
 We have assumed $M=0.1 \sqrt{\theta}$ (dotted curve), $M=0.36 \sqrt{\theta}$ (solid curve),
  $M=1 \sqrt{\theta}$ (dashed curve) and $M=1.8 \sqrt{\theta}$(dot-dashed curve).}
\label{fig1}
\end{figure}

\begin{figure}
\centering
\includegraphics[scale=.4]{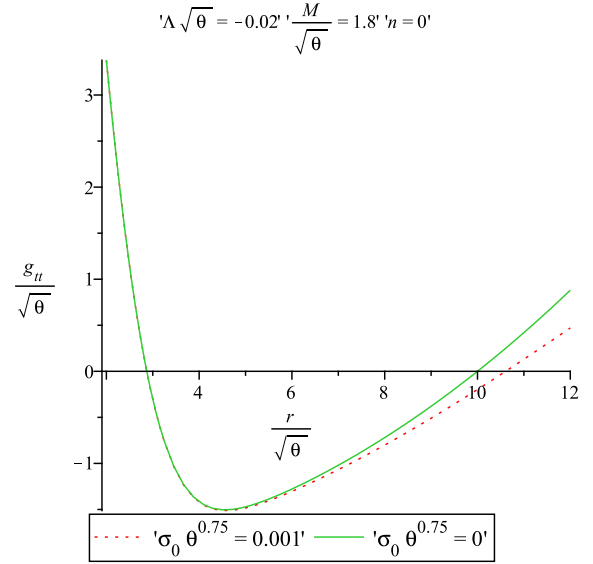}
\caption{$\frac{g_{tt}}{\sqrt{\theta}}$ plotted against
$\frac{r}{\sqrt{\theta}}$ for two different cases: (1) $\sigma_0\theta^{.75}=0.001  $
(dotted curve) and (2) $\sigma_0 =0$( solid curve).}
\label{fig2}
\end{figure}

The condition for the formation of event horizon is given by $g_{tt}(r_h) = 0$, which implies
\begin{equation}
-M+2M e^{-\frac{r_h^{2}}{4\theta}}-\frac{1}{n+2}\left(\frac{4\pi \sigma_0}{n+2}\right)^{2}r_h^{2n+4}-\Lambda r_h^{2} = 0,\label{eq25}
\end{equation}
where $r_h$ is the horizon distance. Though a closed-form solution for $r_h$ cannot be
obtained from the above equation,
 one can express the mass parameter in terms of $r_h$ as
\begin{equation}
M = \frac{\frac{1}{n+2}\left(\frac{4\pi \sigma_0}{n+2}\right)^{2}r_h^{2n+4}+\Lambda r_h^{2}}{-1+2e^{-\frac{r_h^{2}}{4\theta}}}.\label{eq26}
\end{equation}
In Fig.~\ref{fig1}, we have shown the existence of horizons and their corresponding
radii by identifying the intersections of $\frac{g_{tt}}{\sqrt{\theta}}$
with $\frac{r}{\sqrt{\theta}}$. We note that, for a given value of $\Lambda\sqrt{\theta}=-0.02$,
 there exists a minimal mass $M_0 = 0.36\sqrt{\theta}$ below which no black hole exists though
 two distinct horizons exist for $M > 0.36\sqrt{\theta}$. If mass of the black hole increases,
 the distance between the two horizons increases. The two horizons coincide at a minimal mass
  $M = M_0$. In our illustrative example, $r = r_0 = 3.7 \sqrt{\theta}$ is the distance where
   the two horizons coincide, i.e., $r_0$ is the radius of the extremal black hole. The results
    may be summarized as follows:
\begin{itemize}
\item Existence of two horizons for $\frac{M}{\sqrt{\theta}} > M_0$;
\item One horizon corresponding to the extremal black hole with $\frac{M}{\sqrt{\theta}} = M_0$;
\item No horizon for $\frac{M}{\sqrt{\theta}} < M_0$.
\end{itemize}
From Eq.~(\ref{eq26}), it is interesting to note that the lower limit mass $M_0$
 increases when charge is incorporated into the system. Moreover, the distance of
 the outer event horizon increases in the presence of electromagnetic field as shown
  in Fig.~\ref{fig2}.

\begin{figure}
\centering
\includegraphics[scale=.4]{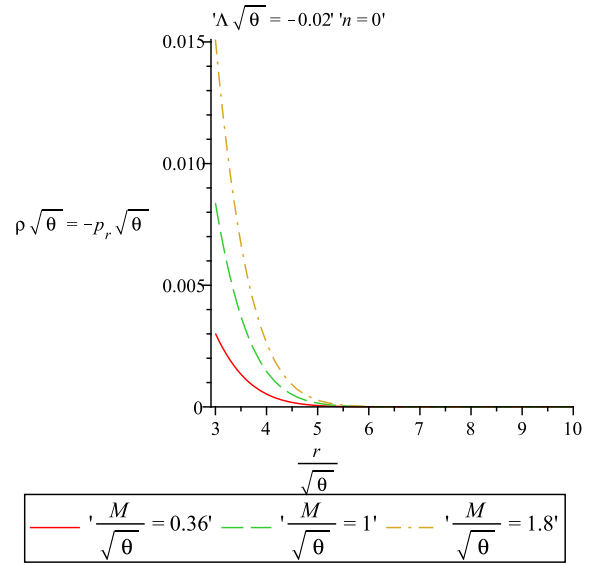}
\caption{$\rho\sqrt{\theta}$ plotted against
$\frac{r}{\sqrt{\theta}}$. We have assumed  $M=0.36
\sqrt{\theta}$ (solid curve), $M=1 \sqrt{\theta}$ (dashed curve)
and $M=1.8 \sqrt{\theta}$(dot-dashed curve).} \label{fig3}
\end{figure}

\begin{figure}
\centering
\includegraphics[scale=.4]{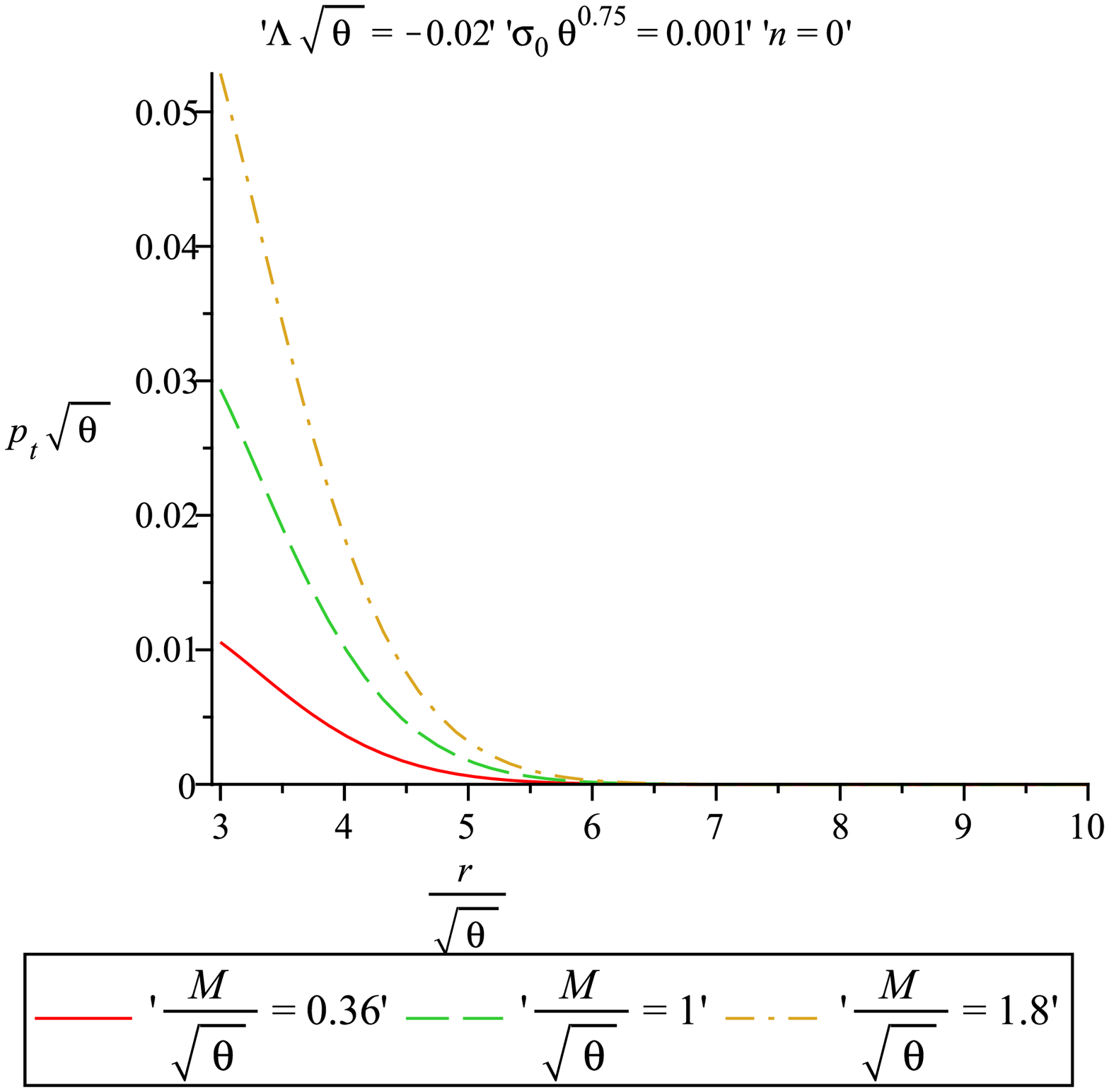}
\caption{ $p_t\sqrt{\theta}$ plotted against
$\frac{r}{\sqrt{\theta}}$.  We have assumed  $M=0.36
\sqrt{\theta}$ (solid curve), $M=1 \sqrt{\theta}$ (dashed curve)
and $M=1.8 \sqrt{\theta}$(dot-dashed curve).} \label{fig4}
\end{figure}

\begin{figure}
\centering
\includegraphics[scale=.4]{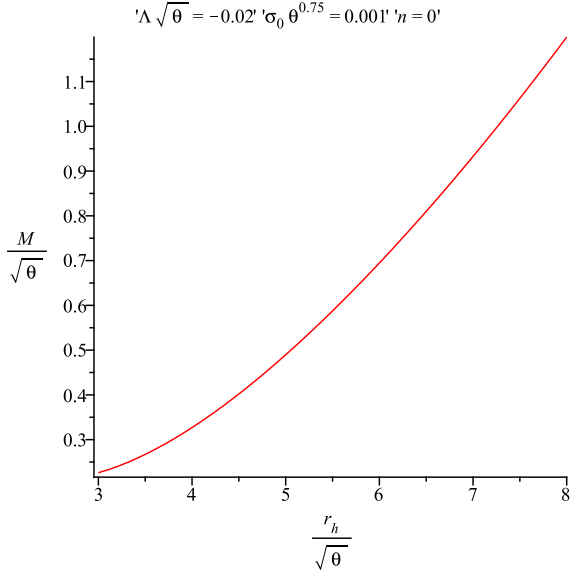}
\caption{Variation of mass $\frac{M}{\sqrt{\theta}}$ with respect to the horizon radius
 $\frac{r_h}{\sqrt{\theta}}$.}
\label{fig5}
\end{figure}

\begin{figure}
\centering
\includegraphics[scale=.4]{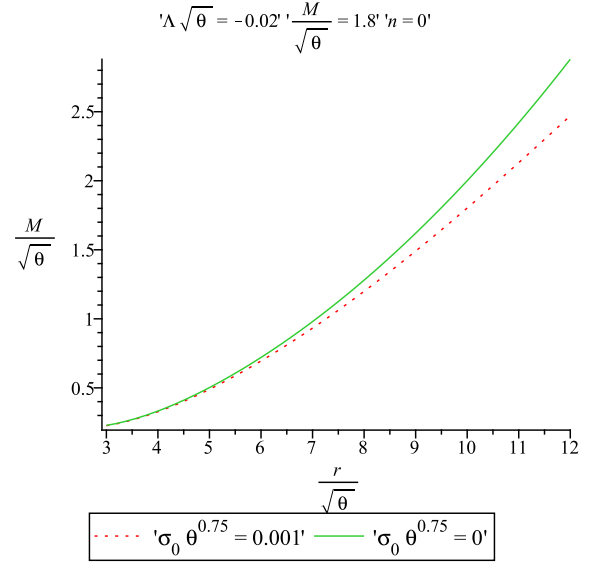}
\caption{Variation of mass $\frac{M}{\sqrt{\theta}}$ with respect
to the horizon radius $\frac{r_h}{\sqrt{\theta}}$ for (1) $\sigma_0\theta^{.75}=0.001$
(dotted curve) and (2) $\sigma_0 =0$( solid curve).}
\label{fig6}
\end{figure}

In Fig.~\ref{fig3}-\ref{fig4}, we have shown behaviour of the density parameter
 ($\rho$)/radial pressure ($p_r\sqrt{\theta}$)
and tangential pressure ($p_t\sqrt{\theta}$), respectively, between the inner and
 outer horizons and outside the outer horizon for different choices of the mass parameter.
  Variation of mass $\frac{M}{\sqrt{\theta}}$ with respect to the horizon radius
   $\frac{r_h}{\sqrt{\theta}}$ has been shown in Fig.~\ref{fig5}. We note a decrease
   in mass $\frac{M}{\sqrt{\theta}}$ within the horizon distance $\frac{r_h}{\sqrt{\theta}}$
    when charge is incorporated into system. This has been shown in Fig.~\ref{fig6}.

\subsection{Thermodynamical  properties }
\begin{figure}
\centering
\includegraphics[scale=.4]{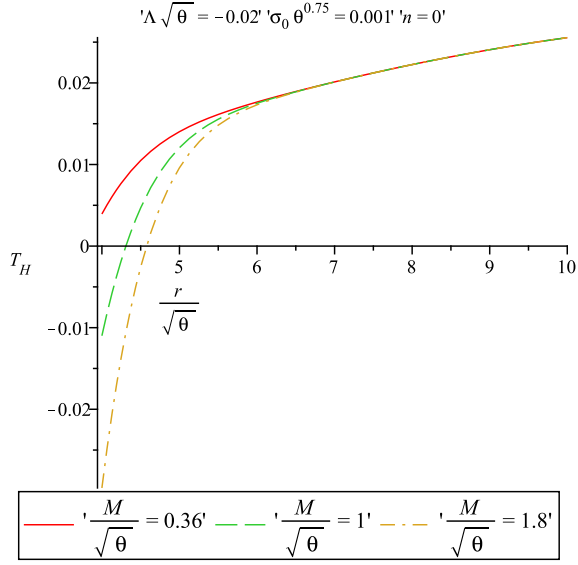}
\caption{Hawking temperature $T_H$ plotted against
$\frac{r}{\sqrt{\theta}}$. We have assumed   $M=0.36
\sqrt{\theta}$ (solid curve), $M=1 \sqrt{\theta}$ (dashed curve)
and $M=1.8 \sqrt{\theta}$(dot-dashed curve).} \label{fig7}
\end{figure}

\begin{figure}
\centering
\includegraphics[scale=.4]{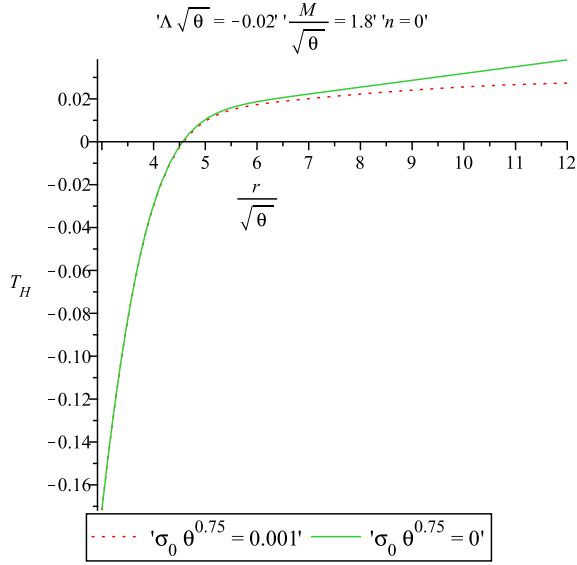}
\caption{The Hawking temperature $T_H$ for (1) $\sigma_0\theta^{.75}=0.001$
 (dotted curve) and (2) $\sigma_0 =0$( solid curve).}
\label{fig8}
\end{figure}

The Hawking temperature
\begin{equation}
T_H = \frac{1}{4\pi}\left(\frac{dg_{tt}}{dr}\right)\sqrt{-g^{tt}g^{rr}}|_{r=r_h},\label{eq27}
\end{equation}
in our model turns out to be
\begin{equation}
T_H = -\frac{r_h}{2\pi}\left[\frac{M}{2\theta}e^{-\frac{r_h^{2}}{4\theta}}
+r_h^{2n+2}\left(\frac{4\pi
\sigma_0}{n+2}\right)^{2}+\Lambda\right].\label{eq28}
\end{equation}
The surface gravity
\begin{equation}
\kappa = \frac{1}{2}\left(\frac{dg_{tt}}{dr}\right)|_{r=r_h},\label{eq29}
\end{equation}
takes the form
\begin{equation}
\kappa = -r_h\left[\frac{M}{2\theta}e^{-\frac{r_h^{2}}{4\theta}}+r_h^{2n+2}\left(\frac{4\pi
\sigma_0}{n+2}\right)^{2}+\Lambda\right],\label{eq30}
\end{equation}
and the Bakenstein-Hawking entropy can be obtained from the equation
\begin{equation}
S = 4\pi r_h.\label{eq31}
\end{equation}
In Fig.~\ref{fig7}, we have shown variation of the Hawking temperature $T_H$ with respect
 to the radial distance $\frac{r}{\sqrt{\theta}}$. Interestingly, the Hawking temperature
 $T_H$ decreases with the inclusion of the charge as shown in Fig.~\ref{fig8}.

\begin{figure}
\centering
\includegraphics[scale=.4]{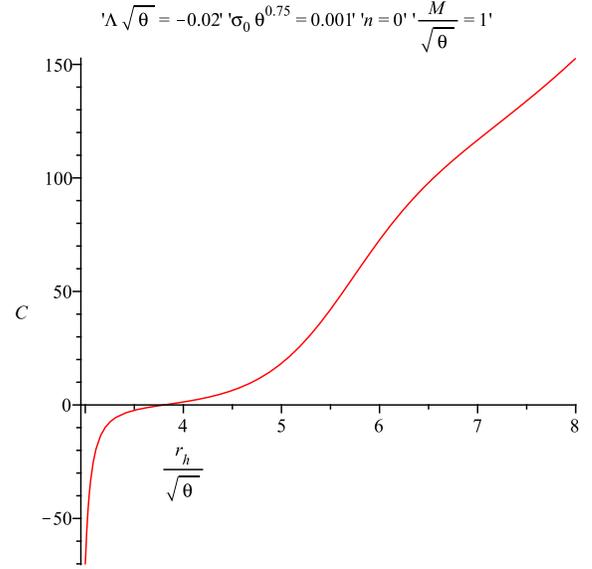}
\caption{Heat capacity $C$ plotted against $\frac{r_h}{\sqrt{\theta}}$.}\label{fig9}
\end{figure}

\begin{figure}
\centering
\includegraphics[scale=.4]{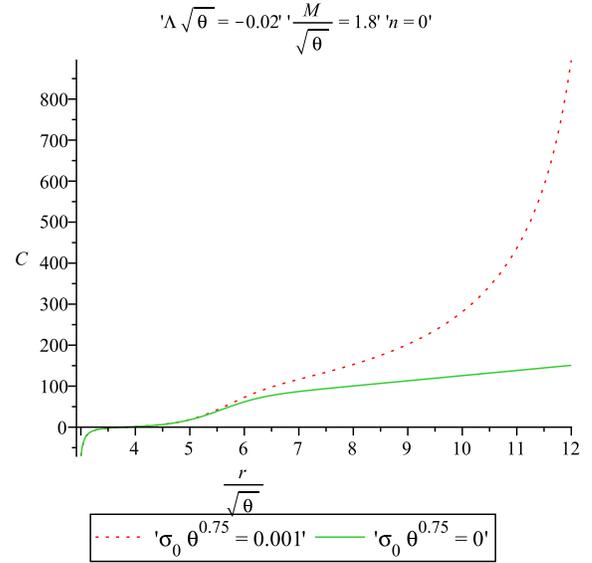}
\caption{Heat capacity $C$ for (1) $\sigma_0\theta^{.75}=0.001$ (dotted curve) and (2)
 $\sigma_0 =0$( solid curve).}
\label{fig10}
\end{figure}

The heat capacity of the charged  BTZ black hole can be obtained from the relation
\[
C = \frac{\partial M(r_h)}{\partial T(r_h)} =
   \frac{\partial M(r_h)}{\partial r_h}\frac{1}
   {\frac{\partial T(r_h)}{\partial r_h}}.
\]
For stability of the black hole, $C$ has to be positive\cite{Liang}.
 In Fig.~\ref{fig9}, the nature of the heat capacity in our model has been shown.
  We note that $C$ vanishes at the extremal event horizon $r_0$. Since $C$ is negative in
   the region $\frac{r_h}{\sqrt{\theta}} < \frac{r_0}{\sqrt{\theta}}$, it is of
    no physical interest. However,
    for $\frac{r_h}{\sqrt{\theta}} > \frac{r_0}{\sqrt{\theta}}$, $C$ is positive
    which indicates that the noncommutative
     geometry inspired charged BTZ black hole is stable in this region. We notice an
     appreciable change in the heat capacity due to the inclusion of charge as shown
     in Fig.~\ref{fig10}.

\subsection{The geodesics}
The geodesics equation
\begin{equation}
\frac{d{x^{\mu}}}{d\tau^{2}}+\Gamma_{\nu \lambda}^{\nu}\frac{dx^{\nu}}{d\tau}
\frac{dx^{\lambda}}{d\tau} = 0,\label{eq32}
\end{equation}
in our construction yields the following results:
\begin{eqnarray}
\frac{1}{f(r)}\left(\frac{dr}{d\tau}\right)^{2} &=& \frac{\Sigma^{2}}{f(r)}-\frac{p^{2}}{r^{2}}+L,\label{eq33}\\
r^{2}\left(\frac{d\phi}{d\tau}\right) &=& p,\label{eq34}\\
\frac{dt}{d\tau} &=& \frac{\Sigma}{f(r)},\label{eq35}
\end{eqnarray}
where, \[f(r)=-M+2M e^{-\frac{r^{2}}{4\theta}}-\frac{1}{n+2}\left(\frac{4\pi
 \sigma_0}{n+2}\right)^{2}r^{2n+4}-\Lambda r^{2}.\]
The constants $\Sigma$ and $p$ represent the energy per unit mass and angular momentum,
 respectively. In Eqs.~(\ref{eq33})-(\ref{eq35}), $\tau$ is the affine parameter and $L$
  is the Lagrangian having values $0$ and $-1$ for massless and massive particles,
   respectively. From Eq.~(\ref{eq33}), we have
\begin{equation}
\frac{1}{2}\left(\frac{dr}{d\tau}\right)^{2} = \frac{1}{2}\left[\Sigma^{2}+f(r)\left
(L-\frac{p^{2}}{r^{2}}\right)\right].\label{eq36}
\end{equation}
Combining Eqs.~(\ref{eq35}) and (\ref{eq36}), we get
\begin{equation}
\left(\frac{dr}{dt}\right)^{2} = \left\{f(r)\right\}^{2}\left[1+\frac{f(r)}{\Sigma^2}\left(L-\frac{p^2}{r^2}\right)\right].\label{eq37}
\end{equation}
Comparing Eq.~(\ref{eq36}) with
$$\frac{1}{2}\left(\frac{dr}{d\tau}\right)^{2} + V_{eff} = 0,$$ we obtain the
 effective potential in the form
\begin{equation}
V_{eff} = -\frac{1}{2}\left[\Sigma^{2}+f(r)\left(L-\frac{p^{2}}{r^{2}}\right)\right].\label{eq38}
\end{equation}

\subsubsection{Null geodesics}
\begin{figure}
\centering
\includegraphics[scale=.4]{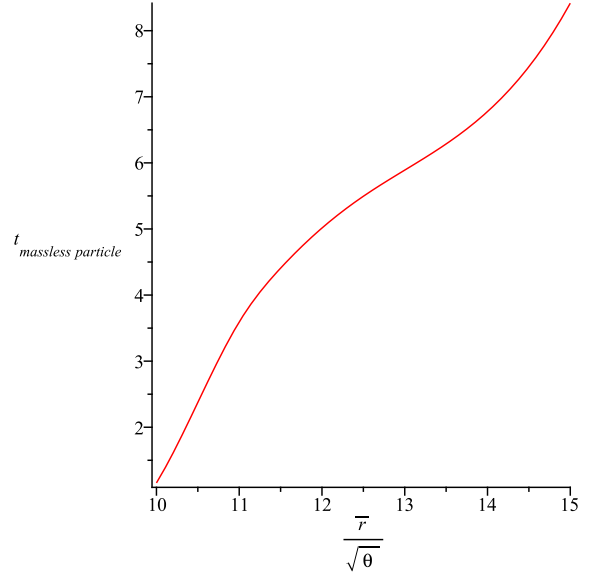}
\caption{Time $\frac{t }{\sqrt{\theta}}$ is plotted against $\frac{r}{\sqrt{\theta}}$.}
\label{fig11}
\end{figure}

For a massless particle, i.e., photon with $L=0$ and $p=0$, we have
\begin{equation}
\left(\frac{dr}{dt}\right)^{2} = \left[-M+2M e^{-\frac{r^{2}}{4\theta}}-\frac{1}{n+2}
\left(\frac{4\pi \sigma_0}{n+2}\right)^{2}r^{2n+4}-\Lambda r^{2}\right]^{2},\label{eq39}
\end{equation}
which in integral form can be expressed as
\begin{equation}
\pm t=\int_{\frac{r_h}{\sqrt{\theta}}}^{\frac{r^{*}}{\sqrt{\theta}}}\frac{dr}{-M+2M
 e^{-\frac{r^{2}}{4\theta}}-\frac{1}{n+2}\left(\frac{4\pi \sigma_0}{n+2}\right)^{2}r^{2n+4}
 -\Lambda r^{2}}.\label{eq40}
\end{equation}
Eq.~(\ref{eq40}) is not integrable, however, one can obtain the values of $t$ for
different choices of ${\frac{r^{*}}{\sqrt{\theta}}}$ by numerical procedures.
This has been shown in Table$~1$ and a graphical representation of it is given in
 Fig.~\ref{fig11}.
\begin{table}
\caption{Values of $\frac{t}{\sqrt{\theta}}$  for different
$\frac{r^{*}}{\sqrt{\theta}}$. (We have assumed,
$\frac{r_h}{\sqrt{\theta}} = 10$,
$\frac{M}{\sqrt{\theta}} = 1.8$.) }
{\begin{tabular}{@{}|c|c|@{}} \toprule $\frac{r^{*}}{\sqrt{\theta}}$ & $\frac{t}
{\sqrt{\theta}}$ \\
\colrule
10 & 0\\
10.5 & 2.3753  \\
11 & 3.5934 \\
11.5 & 4.4077  \\
12 & 5.01449  \\
12.5 & 5.4950 \\
13 & 5.8907\\
\botrule
\end{tabular}}
\end{table}

\begin{figure}
\centering
\includegraphics[scale=.4]{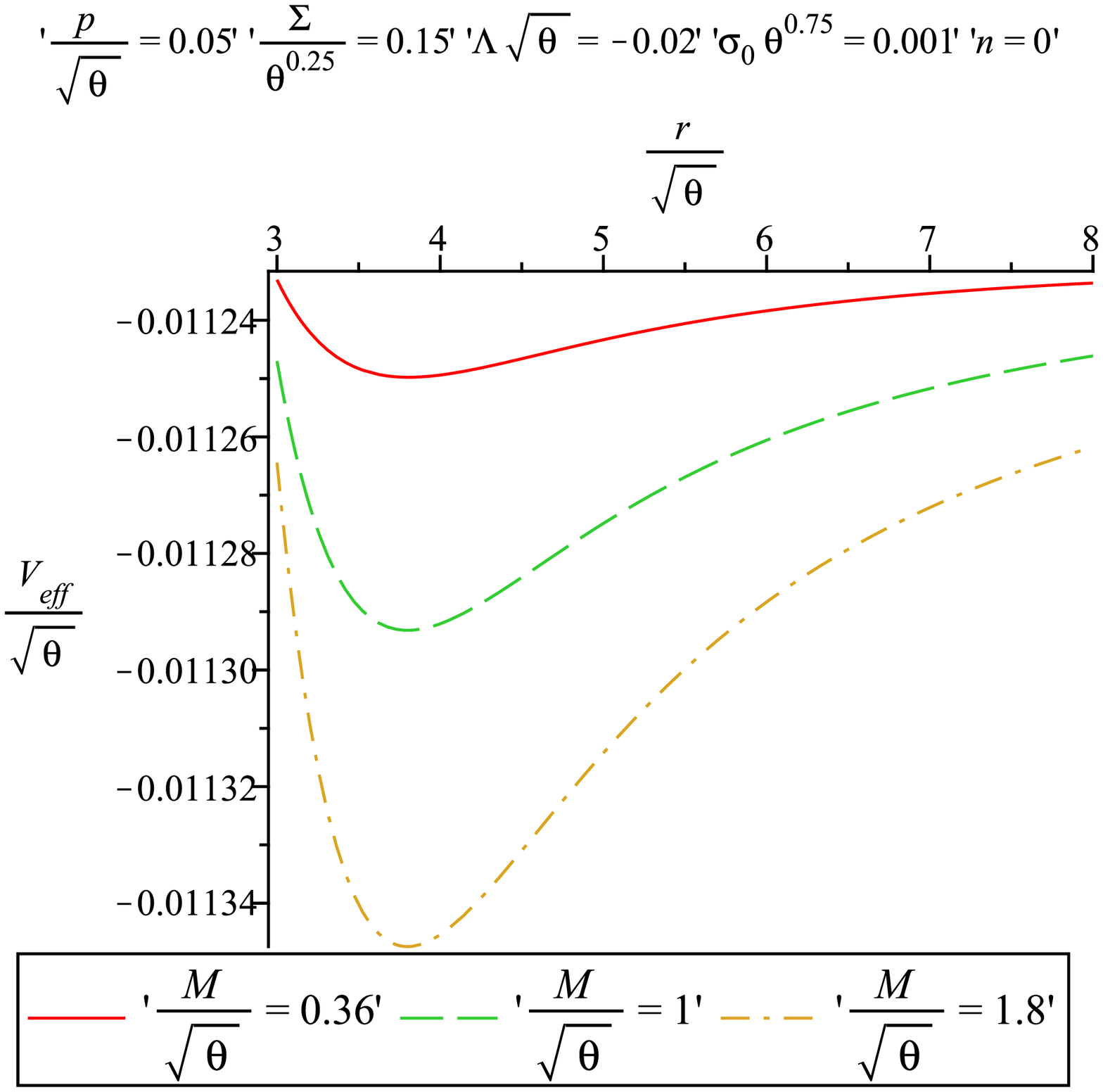}
\caption{Effective potential $\frac{V_{eff}}{\sqrt{\theta}}$ for
massless particles.
Here, $M=0.36 \sqrt{\theta}$ (solid curve), $M=1 \sqrt{\theta}$
(dashed curve) and $M=1.8 \sqrt{\theta}$ (dot-dashed curve).}
\label{fig12}
\end{figure}

The effective potential of the system is obtained as
\begin{eqnarray}
V_{eff} &=& -\frac{\Sigma^{2}}{2}+ \frac{p^{2}}{2r^{2}}\left[-M+2Me^{-\frac{r^{2}}{4\theta}}-\right.\nonumber\\
&&\left.\frac{1}{n+2}\left(\frac{4\pi
\sigma_0}{n+2}\right)^{2}r^{2n+4}-\Lambda r^{2}\right],\label{eq41}
\end{eqnarray}
which for $p=0$ reduces to
$$V_{eff} = -\frac{\Sigma^{2}}{2}.$$
The nature of $V_{eff}$ for massless particles has been shown in Fig.~\ref{fig12}.

\subsubsection{Time-like geodesics}
\begin{figure}
\centering
\includegraphics[scale=.4]{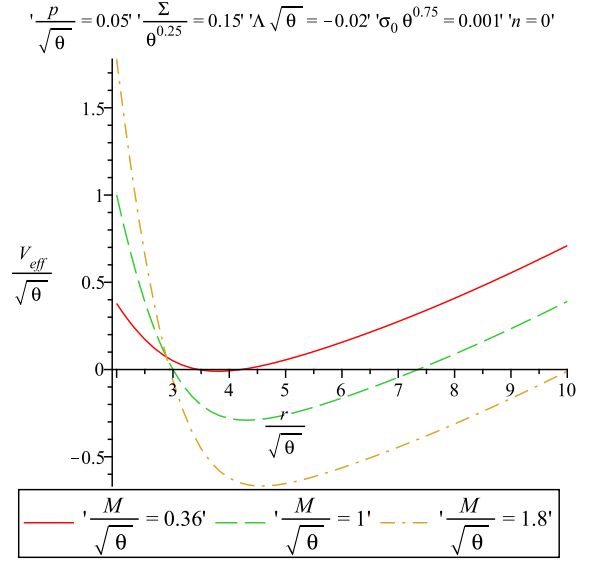}
\caption{Effective potential $\frac{V_{eff}}{\sqrt{\theta}}$ for
massive  particles.
Here,  $M=0.36 \sqrt{\theta}$ (solid curve), $M=1 \sqrt{\theta}$
(dashed curve) and $M=1.8 \sqrt{\theta}$ (dot-dashed curve).}
\label{fig13}
\end{figure}

For massive particles with $L = -1$ and $p=0$, we have
\begin{equation}
\left(\frac{dr}{dt}\right)^{2}=\left(\frac{f(r)}{\Sigma}\right)^{2}(\Sigma^{2}-f(r)),\label{eq42}
\end{equation}
which in integral form can be written as
\begin{equation}
\pm t= \Sigma \int\frac{dr}{f(r)\sqrt{\Sigma^{2}-f(r)}}.\label{eq43}
\end{equation}
For a time-like particle with $p \neq 0$, the effective potential $V_{eff}$ gets the form
\begin{equation}
V_{eff} = -\frac{1}{2}\Sigma^{2}+\frac{f(r)}{2}\left(1+\frac{p^{2}}{r^{2}}\right).\label{eq44}
\end{equation}
The nature of effective potential for massive particles has been shown in Fig.~\ref{fig13}.
 The shape of the effective potential indicates the possibility of stable circular orbits for
  massive particles around the black hole.

\section{Test particles}
Let us now analyze the motion of a test particle around noncommutative charged black
hole using Hamilton-Jacobi(H-J) approach. Let $m$ and $e$ denote the respective mass
 and charge of the test particle. The Hamilton-Jacobi equation for the test particle
 has the standard form
\begin{equation}
g^{ik} \left( \frac{\partial S}{\partial x^i} + e A_i \right) \left(
\frac{\partial S}{\partial x^k} + e A_k \right) +m^2  = 0,\label{eq45}
\end{equation}
where,  $g_{ik}$ is  the background space and $S(t,r, \phi)$ is the Hamilton's
 characteristic function. The gauge potential $A_i$ is given by
\[ A_i = \frac{Q dt}{r},\]
where $Q$ is the total charge of the black hole. For the background space-time (\ref{eq1}),
the H-J equation gets the form
\begin{eqnarray}
-\frac{1}{f} \left( \frac{\partial S}{\partial t} + \frac{e Q}{r}
\right)^2 + f \left( \frac{\partial S}{\partial r} \right)^2 +
\frac{1}{r^2}\left( \frac{\partial S}{\partial \phi}\right)^2
  + m^2 = 0.\label{eq46}
\end{eqnarray}

To solve the above partial differential equation, we use the method of separation of variables
and choose the H-J function $S$ as
\[ S(t,r,\phi) = -E t + S_1(r)  + p \phi, \]
where, $E$ is the energy and $p$ is the angular momentum of the particle. The radial velocity
 of the particle is then obtained as
\begin{equation}
\frac{ dr}{dt} = f^2\left(E - \frac{e Q}{r^2}\right)^{-1}
\sqrt{ \frac{1}{f^2} \left(E-\frac{e Q}{r} \right)^2 -
\frac{m^2}{f} - \frac{p^2}{fr^2}},\label{eq47}
\end{equation}
  Now, the turning points of the trajectory are determined from the requirement $dr/dt = 0$,
  which yields
\begin{equation}
\left(E-\frac{e Q}{r} \right)^2 - m^2f - \frac{p^2}{r^2}f = 0.\label{eq48}
\end{equation}
Consequently, the potential curve can be obtained from the relation
\begin{equation}
V(r) \equiv \frac{E}{m} = \frac{eQ}{m r} + \sqrt{f} \left( 1+\frac{p^2}{m^2r^2} \right)^{1/2}.\label{eq49}
\end{equation}
For a stationary system, $V(r)$ must have an extremal value, i.e., $dV/dr = 0$ which implies
\begin{equation}
\frac{e Q}{m r^2}\sqrt{f}\left( 1+ \frac{p^2}{m^2r^2} \right)^{1/2} = \frac{1}{2} \left(1 + \frac{p^2}{m^2r^2} \right) f^\prime  -
f\frac{p^2}{m^2r^3}.\label{eq50}
\end{equation}

\subsection{Test particle in static equilibrium}
\begin{figure}
\centering
\includegraphics[scale=.4]{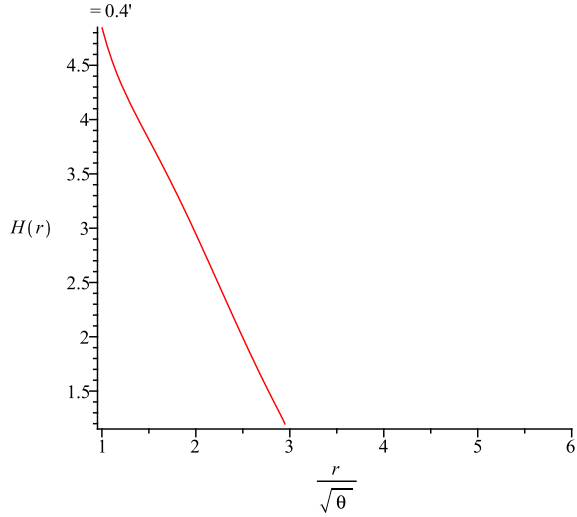}
\caption{$H(r)$ is  plotted against $\frac{r}{\sqrt{\theta}}$.}
\label{fig14}
\end{figure}

In static equilibrium ($p = 0$), from Eq.~(\ref{eq50}), the value of $r$ at which potential curve will be an extremal can be obtained from the requirement
\begin{widetext}
\begin{equation}
H(r) \equiv \frac{e Q}{mr^2}\left[-M+2M e^{-\frac{r^{2}}{4\theta}}-\frac{1}{n+2}\left(\frac{4\pi
\sigma_0}{n+2}\right)^{2}r^{2n+4}-\Lambda r^{2}\right]^{\frac{1}{2}}+\frac{Mr}{2\theta}e^{-\frac{r^{2}}{4\theta}}
+r^{2n+3}\left(\frac{4\pi \sigma_0}{n+2}  \right)^{2}+r\Lambda = 0.\label{eq51}
\end{equation}
\end{widetext}
A graphical representation of $H(r)$, as shown in Fig.~\ref{fig14}, indicates that
 there does not exist any $r$ for which $H(r) = 0$. This implies that the test particle
  can not be trapped by the black hole. In other words, the black exerts no gravitation
  force on the test particle in static equilibrium situation.

\subsection{Test particle in non-static equilibrium}
For a text particle in non-static equilibrium ($p \neq 0$), we consider two possibilities:
\subsubsection{Uncharged test particle $(e=0)$}
\begin{figure}
\centering
\includegraphics[scale=.4]{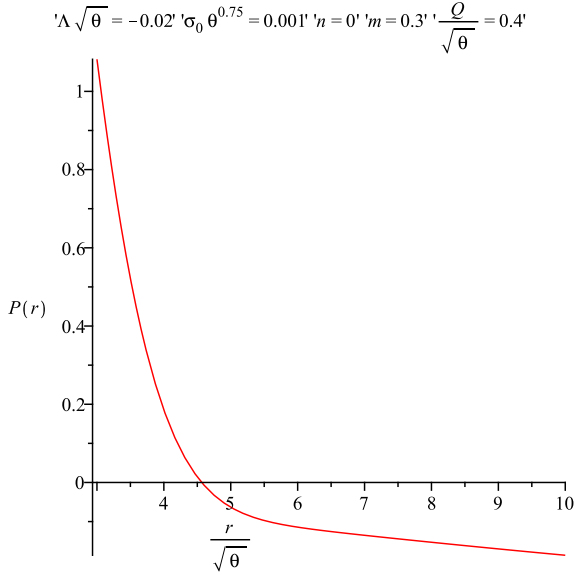}
\caption{$P(r)$ is  plotted against $\frac{r}{\sqrt{\theta}}$.}
\label{fig15}
\end{figure}

In this case, the value of $r$ at which potential curve will be an extremal can be obtained
from the requirement
\begin{widetext}
\begin{equation}
P(r)\equiv \left(1+\frac{p^{2}}{r^{2}m^{2}}\right)\left[\frac{Mr}{2\theta}e^{-\frac{r^{2}}{4\theta}}
+r^{2n+3}\left(\frac{4\pi \sigma_0}{n+2}  \right)^{2}+r\Lambda \right]-\frac{p^{2}}{m^{2}r^{3}}
\left[ -M+2M e^{-\frac{r^{2}}{4\theta}}-\frac{1}{n+2}\left(\frac{4\pi \sigma_0}{n+2}\right)^{2}r^{2n+4}-\Lambda r^{2}  \right] = 0.\label{eq52}
\end{equation}
\end{widetext}
In Fig.~\ref{fig15}, we note that $P(r) =0$ at some finite value of $r$. Therefore,
the particle can be trapped by the black hole. In other words, the black hole exerts
 gravitation force on the uncharged test particle in non-static equilibrium.

\subsubsection{Test particle with charge $( e \neq 0 )$}
\begin{figure}
\centering
\includegraphics[scale=0.4]{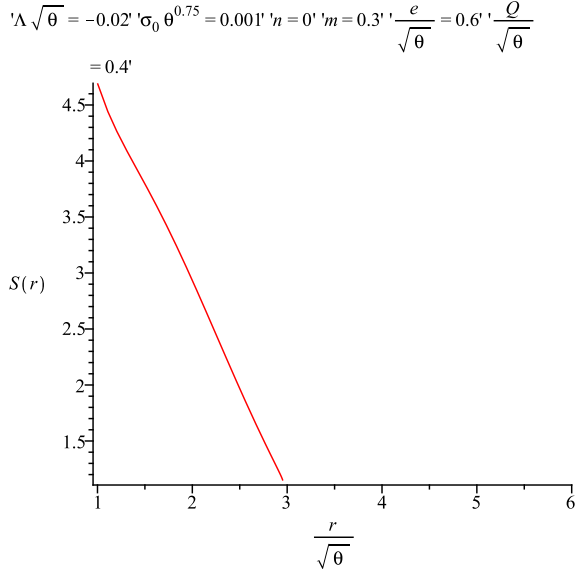}
\caption{$S(r)$ is  plotted against $\frac{r}{\sqrt{\theta}}$.}
\label{fig16}
\end{figure}

For a charged particle, the value of $r$ at which potential curve will be an extremal can be obtained from the requirement
\begin{widetext}
\begin{eqnarray}
 S(r) \equiv
\frac{eQ}{mr^2}\left(1+\frac{p^{2}}{m^{2}r^{2}}\right)^{\frac{1}{2}}\left[-M+2M
e^{-\frac{r^{2}}{4\theta}}-\frac{1}{n+2}\left(\frac{4\pi
\sigma_0}{n+2}\right)^{2}r^{2n+4}-\Lambda r^{2}
\right]^{\frac{1}{2}}\nonumber\\
+\left(1+\frac{p^{2}}{r^{2}m^{2}}\right)\left[\frac{Mr}{2\theta}e^{-\frac{r^{2}}{4\theta}}+r^{2n+3}\left(\frac{4\pi
\sigma_0}{n+2}  \right)^{2}+r\Lambda\right]
-\frac{p^{2}}{m^{2}r^{3}}\left[-M+2M e^{-\frac{r^{2}}{4\theta}}-\frac{1}{n+2}\left(\frac{4\pi \sigma_0}{n+2}\right)^{2}r^{2n+4}-\Lambda r^{2}
 \right] = 0.\label{eq53}
\end{eqnarray}
\end{widetext}
Fig.~\ref{fig16} shows that $S(r) \neq 0$ for any finite value of $r$.
This implies that the black exerts no gravitation force on the
charged test particle in non-static equilibrium.

The results clearly show that a non-static charged test particle behaves differently
 from its neutral counterpart in the gravitational field of a charged black hole.
 The motion of a neutral particle is solely governed by the gravitational field of
  the black hole whereas the motion of a charged particle in the gravitational field
   of a charged black hole also experiences electromagnetic force. Therefore, while a
    neutral particle moves along the geodesic and gets trapped, a charged test particle
     in motion may not be trapped by the black hole.

\section{Discussions}
In this work, we have constructed a charged BTZ-like black hole solution by adopting the
formalism of noncommutative geometry.
The uncharged limit of this solution reported in Ref.~\cite{Rahaman4} can be regained
simply by putting $\sigma =
0$. The noncommutative effects on the physical quantities are found to be similar to
 the properties discussed in
Ref.~\cite{Rahaman4}. To investigate the impact of electromagnetic field we have
 considered two cases ($\sigma =0$ and $\sigma \neq 0$)
and compared the results graphically. Our results show that the limiting mass $M_0$
 below which no black hole can exist depends on its associated
 charge. In fact, the value of the limiting mass increases as charge
is incorporated into the system. The location of the event horizon,
 Hawking temperature and heat capacity also do get modified in the
  presence of the electromagnetic field. Particle trajectories around
   the black hole are also affected by the inclusion charge. In particular,
    we note that the motion of a charged particle around a charged black hole
    is quite different from the motion of a neutral test particle. We hope that
     our results will contribute towards getting a better understanding of the
     effects of noncommutative geometry on the structure and properties of black
     holes. In addition, the results may be useful to understand how the presence
     of electromagnetic field can change a number of astrophysical phenomena such
      as properties of accretion
 disks around charged black holes and Hawking radiation.

\begin{acknowledgments}
RS and FR gratefully acknowledge support from the
Inter-University Centre for Astronomy and Astrophysics (IUCAA),
Pune, India, where a part of this work was carried out under its
Visiting Research Associateship Programme. We are very grateful
to an anonymous referee for his/her insightful comments that have
led to significant improvements, particularly on the
interpretational aspects.
\end{acknowledgments}

\end{document}